\documentclass[aps,prc,superscriptaddress,showpacs,nofootinbib,floatfix,twocolumn]{revtex4}

\usepackage{graphicx}   

\def\pt{$p_{\rm T}$}

\def\zt{$z_{\rm T}$}

\def\raa{$R_{\rm AA}$}
\def\iaa{$I_{\rm AA}$}
\def\auau{Au+Au}
\def\pp{$p$+$p$}
\def\piz{$\pi^{0}$}
\def\rg{$R_{\gamma}$}

\def\ptg{$p_{\rm T}^{\gamma}\,$}
\def\pth{$p_{\rm T}^{\rm h}\,$}
\def\ptp{$p_{\rm T}^{\pi}\,$}
\def\ptgrange{$< p_{\rm T}^{\gamma} <$}
\def\pthrange{$< p_{\rm T}^{\rm h} <$}

\def\dphi{$\Delta\phi$}
\def\vtwo{$v_2$}
\def\mean#1{\left<#1\right>}

\begin{document}

\title{Photon-Hadron Jet Correlations in $p$+$p$ and Au+Au 
       Collisions at $\sqrt{s_{NN}}$=200 GeV}

\newcommand{\abilene}{Abilene Christian University, Abilene, TX 79699, U.S.}
\newcommand{\acadsin}{Institute of Physics, Academia Sinica, Taipei 11529, Taiwan}
\newcommand{\banaras}{Department of Physics, Banaras Hindu University, Varanasi 221005, India}
\newcommand{\barc}{Bhabha Atomic Research Centre, Bombay 400 085, India}
\newcommand{\bnlcoll}{Collider-Accelerator Department, Brookhaven National Laboratory, Upton, NY 11973-5000, U.S.}
\newcommand{\bnlphys}{Physics Department, Brookhaven National Laboratory, Upton, NY 11973-5000, U.S.}
\newcommand{\caucr}{University of California - Riverside, Riverside, CA 92521, U.S.}
\newcommand{\charlesczech}{Charles University, Ovocn\'{y} trh 5, Praha 1, 116 36, Prague, Czech Republic}
\newcommand{\ciae}{China Institute of Atomic Energy (CIAE), Beijing, People's Republic of China}
\newcommand{\cns}{Center for Nuclear Study, Graduate School of Science, University of Tokyo, 7-3-1 Hongo, Bunkyo, Tokyo 113-0033, Japan}
\newcommand{\colorado}{University of Colorado, Boulder, CO 80309, U.S.}
\newcommand{\columbia}{Columbia University, New York, NY 10027 and Nevis Laboratories, Irvington, NY 10533, U.S.}
\newcommand{\czechtech}{Czech Technical University, Zikova 4, 166 36 Prague 6, Czech Republic}
\newcommand{\dapnia}{Dapnia, CEA Saclay, F-91191, Gif-sur-Yvette, France}
\newcommand{\debrecen}{Debrecen University, H-4010 Debrecen, Egyetem t{\'e}r 1, Hungary}
\newcommand{\elte}{ELTE, E{\"o}tv{\"o}s Lor{\'a}nd University, H - 1117 Budapest, P{\'a}zm{\'a}ny P. s. 1/A, Hungary}
\newcommand{\fit}{Florida Institute of Technology, Melbourne, FL 32901, U.S.}
\newcommand{\fsu}{Florida State University, Tallahassee, FL 32306, U.S.}
\newcommand{\gsu}{Georgia State University, Atlanta, GA 30303, U.S.}
\newcommand{\hiroshima}{Hiroshima University, Kagamiyama, Higashi-Hiroshima 739-8526, Japan}
\newcommand{\ihepprot}{IHEP Protvino, State Research Center of Russian Federation, Institute for High Energy Physics, Protvino, 142281, Russia}
\newcommand{\illuiuc}{University of Illinois at Urbana-Champaign, Urbana, IL 61801, U.S.}
\newcommand{\instpasczech}{Institute of Physics, Academy of Sciences of the Czech Republic, Na Slovance 2, 182 21 Prague 8, Czech Republic}
\newcommand{\isu}{Iowa State University, Ames, IA 50011, U.S.}
\newcommand{\jinrdubna}{Joint Institute for Nuclear Research, 141980 Dubna, Moscow Region, Russia}
\newcommand{\kaeri}{KAERI, Cyclotron Application Laboratory, Seoul, Korea}
\newcommand{\kek}{KEK, High Energy Accelerator Research Organization, Tsukuba, Ibaraki 305-0801, Japan}
\newcommand{\kfki}{KFKI Research Institute for Particle and Nuclear Physics of the Hungarian Academy of Sciences (MTA KFKI RMKI), H-1525 Budapest 114, POBox 49, Budapest, Hungary}
\newcommand{\korea}{Korea University, Seoul, 136-701, Korea}
\newcommand{\kurchatov}{Russian Research Center ``Kurchatov Institute", Moscow, Russia}
\newcommand{\kyoto}{Kyoto University, Kyoto 606-8502, Japan}
\newcommand{\labllr}{Laboratoire Leprince-Ringuet, Ecole Polytechnique, CNRS-IN2P3, Route de Saclay, F-91128, Palaiseau, France}
\newcommand{\lawllnl}{Lawrence Livermore National Laboratory, Livermore, CA 94550, U.S.}
\newcommand{\losalamos}{Los Alamos National Laboratory, Los Alamos, NM 87545, U.S.}
\newcommand{\lpc}{LPC, Universit{\'e} Blaise Pascal, CNRS-IN2P3, Clermont-Fd, 63177 Aubiere Cedex, France}
\newcommand{\lund}{Department of Physics, Lund University, Box 118, SE-221 00 Lund, Sweden}
\newcommand{\mass}{Department of Physics, University of Massachusetts, Amherst, MA 01003-9337, U.S. }
\newcommand{\muenster}{Institut f\"ur Kernphysik, University of Muenster, D-48149 Muenster, Germany}
\newcommand{\muhlenberg}{Muhlenberg College, Allentown, PA 18104-5586, U.S.}
\newcommand{\myongji}{Myongji University, Yongin, Kyonggido 449-728, Korea}
\newcommand{\nagasaki}{Nagasaki Institute of Applied Science, Nagasaki-shi, Nagasaki 851-0193, Japan}
\newcommand{\newmex}{University of New Mexico, Albuquerque, NM 87131, U.S. }
\newcommand{\nmsu}{New Mexico State University, Las Cruces, NM 88003, U.S.}
\newcommand{\ornl}{Oak Ridge National Laboratory, Oak Ridge, TN 37831, U.S.}
\newcommand{\orsay}{IPN-Orsay, Universite Paris Sud, CNRS-IN2P3, BP1, F-91406, Orsay, France}
\newcommand{\peking}{Peking University, Beijing, People's Republic of China}
\newcommand{\pnpi}{PNPI, Petersburg Nuclear Physics Institute, Gatchina, Leningrad region, 188300, Russia}
\newcommand{\riken}{RIKEN Nishina Center for Accelerator-Based Science, Wako, Saitama 351-0198, JAPAN}
\newcommand{\rikjrbrc}{RIKEN BNL Research Center, Brookhaven National Laboratory, Upton, NY 11973-5000, U.S.}
\newcommand{\rikkyo}{Physics Department, Rikkyo University, 3-34-1 Nishi-Ikebukuro, Toshima, Tokyo 171-8501, Japan}
\newcommand{\saispbstu}{Saint Petersburg State Polytechnic University, St. Petersburg, Russia}
\newcommand{\saopaulo}{Universidade de S{\~a}o Paulo, Instituto de F\'{\i}sica, Caixa Postal 66318, S{\~a}o Paulo CEP05315-970, Brazil}
\newcommand{\seoulnat}{System Electronics Laboratory, Seoul National University, Seoul, Korea}
\newcommand{\stonybrkc}{Chemistry Department, Stony Brook University, Stony Brook, SUNY, NY 11794-3400, U.S.}
\newcommand{\stonycrkp}{Department of Physics and Astronomy, Stony Brook University, SUNY, Stony Brook, NY 11794, U.S.}
\newcommand{\subatech}{SUBATECH (Ecole des Mines de Nantes, CNRS-IN2P3, Universit{\'e} de Nantes) BP 20722 - 44307, Nantes, France}
\newcommand{\tenn}{University of Tennessee, Knoxville, TN 37996, U.S.}
\newcommand{\titech}{Department of Physics, Tokyo Institute of Technology, Oh-okayama, Meguro, Tokyo 152-8551, Japan}
\newcommand{\tsukuba}{Institute of Physics, University of Tsukuba, Tsukuba, Ibaraki 305, Japan}
\newcommand{\vandy}{Vanderbilt University, Nashville, TN 37235, U.S.}
\newcommand{\waseda}{Waseda University, Advanced Research Institute for Science and Engineering, 17 Kikui-cho, Shinjuku-ku, Tokyo 162-0044, Japan}
\newcommand{\weizmann}{Weizmann Institute, Rehovot 76100, Israel}
\newcommand{\yonsei}{Yonsei University, IPAP, Seoul 120-749, Korea}
\affiliation{\abilene}
\affiliation{\acadsin}
\affiliation{\banaras}
\affiliation{\barc}
\affiliation{\bnlcoll}
\affiliation{\bnlphys}
\affiliation{\caucr}
\affiliation{\charlesczech}
\affiliation{\ciae}
\affiliation{\cns}
\affiliation{\colorado}
\affiliation{\columbia}
\affiliation{\czechtech}
\affiliation{\dapnia}
\affiliation{\debrecen}
\affiliation{\elte}
\affiliation{\fit}
\affiliation{\fsu}
\affiliation{\gsu}
\affiliation{\hiroshima}
\affiliation{\ihepprot}
\affiliation{\illuiuc}
\affiliation{\instpasczech}
\affiliation{\isu}
\affiliation{\jinrdubna}
\affiliation{\kaeri}
\affiliation{\kek}
\affiliation{\kfki}
\affiliation{\korea}
\affiliation{\kurchatov}
\affiliation{\kyoto}
\affiliation{\labllr}
\affiliation{\lawllnl}
\affiliation{\losalamos}
\affiliation{\lpc}
\affiliation{\lund}
\affiliation{\mass}
\affiliation{\muenster}
\affiliation{\muhlenberg}
\affiliation{\myongji}
\affiliation{\nagasaki}
\affiliation{\newmex}
\affiliation{\nmsu}
\affiliation{\ornl}
\affiliation{\orsay}
\affiliation{\peking}
\affiliation{\pnpi}
\affiliation{\riken}
\affiliation{\rikjrbrc}
\affiliation{\rikkyo}
\affiliation{\saispbstu}
\affiliation{\saopaulo}
\affiliation{\seoulnat}
\affiliation{\stonybrkc}
\affiliation{\stonycrkp}
\affiliation{\subatech}
\affiliation{\tenn}
\affiliation{\titech}
\affiliation{\tsukuba}
\affiliation{\vandy}
\affiliation{\waseda}
\affiliation{\weizmann}
\affiliation{\yonsei}
\author{A.~Adare} \affiliation{\colorado}
\author{S.~Afanasiev} \affiliation{\jinrdubna}
\author{C.~Aidala} \affiliation{\columbia} \affiliation{\mass}
\author{N.N.~Ajitanand} \affiliation{\stonybrkc}
\author{Y.~Akiba} \affiliation{\riken} \affiliation{\rikjrbrc}
\author{H.~Al-Bataineh} \affiliation{\nmsu}
\author{J.~Alexander} \affiliation{\stonybrkc}
\author{A.~Al-Jamel} \affiliation{\nmsu}
\author{K.~Aoki} \affiliation{\kyoto} \affiliation{\riken}
\author{L.~Aphecetche} \affiliation{\subatech}
% ...
\collaboration{PHENIX Collaboration} \noaffiliation

\date{\today}

\begin{abstract}

We report the observation at the Relativistic Heavy Ion Collider (RHIC)
of suppression of back-to-back correlations in the direct 
photon+jet channel in Au+Au relative to $p$+$p$ collisions. 
Two-particle correlations of direct photon triggers
with associated hadrons are obtained by statistical subtraction of
the decay photon-hadron ($\gamma$-$h$) background.  
The initial momentum of the
away-side parton is tightly constrained, because the parton-photon pair
exactly balance in momentum at leading order in perturbative quantum
chromodynamics (pQCD), making such
correlations a powerful probe of the in-medium parton energy loss.
The away-side nuclear suppression factor, \iaa, in central Au+Au
collisions, is $0.32 \pm 0.12^{\rm stat} \pm 0.09^{\rm syst}$ for hadrons 
of 3 \pthrange\ 5 in coincidence with photons of 5 \ptgrange\ 15 
GeV/c.  The suppression is comparable to that observed for high-\pt\ single
hadrons and dihadrons.  The direct photon associated yields in $p$+$p$
collisions scale approximately with the momentum balance,
\zt\ $\equiv$ \pth\ / \ptg, as expected for a measure of the away-side
parton fragmentation function. We compare to Au+Au collisions for 
which the momentum balance dependence of the nuclear modification
should be sensitive to the path-length dependence of parton energy loss.

\end{abstract}


\maketitle



\section{Introduction}

Experimental results from RHIC  
have established the formation of hot and dense matter of a 
fundamentally new nature in relativistic heavy-ion collisions at 
$\sqrt{s_{NN}}=200$ GeV~\cite{whitepaper}.  Energy loss in this 
dense nuclear matter by color-charged, hard (E $\gtrsim 2$ GeV) 
partons, and the jets into which they fragment, is generally 
accepted to be the mechanism responsible for the suppression of the 
high-\pt\ hadron yields observed in central A+A 
collisions~\cite{ppg080,Muller:2006ee}.  In the large multiplicity 
environment of heavy-ion collisions, two-particle correlations are 
often used to study jet modification and to infer properties of the 
medium. For example, high-\pt\ azimuthal dihadron correlations 
demonstrate that the degree of dijet away-side suppression depends 
on the \pt\ of the ``trigger'' and ``associated'' hadrons.  
At moderate \pt\ ($\gtrsim3$ GeV/c), the jet properties measured 
through two-particle correlations demonstrate novel features such 
as shape modifications which are thought to be a manifestation of 
the response of medium to the energy deposited by the attenuated 
parton~\cite{ppg083}.

Di-hadron measurements of dijet pairs provide an ambiguous 
measurement of the energy loss of the away-side parton. The trigger 
hadron is a product of parton fragmentation and therefore it is not 
possible to determine, event-by-event, whether the near-side parton 
has itself lost energy. Given the steeply falling jet spectrum, the 
sample of hard scatterings is biased towards configurations in 
which the parton loses little energy.  In particular, it is 
believed that hadron measurements are subject to a ``surface bias'' 
in which the hard scatterings sampled are likely to occur at the 
periphery of the overlap zone \cite{surfbias1,surfbias2}.  The 
away-side parton then is more likely to traverse a maximal 
path-length through the medium. For a sufficiently opaque medium, 
the attenuation of the parton may be nearly total, in which case 
the sensitivity to the average path-length is reduced 
~\cite{Eskola:2004cr}. Back-to-back, high-\pt\ hadron pairs may 
originate preferentially from configurations in which the outgoing 
parton trajectories are tangential to the surface of the overlap 
zone~\cite{Loizides:2006zz}.
On the other hand, dihadron pairs may also originate from
vertices deep in the collision zone if a parton has a finite
probability to ``punch-through'' or pass through the medium without
interaction~\cite{Renk:2006pk}.
Calculations of the relative importance of these two 
mechanisms depend both on the model of parton energy loss employed 
and the density profile of the 
medium~\cite{surfbias2,ZOWW1,jjia_drees}.

Direct photon-jet pairs offer two major advantages in studying 
energy loss as compared to dijets because of the nature of the 
photon. First, in contrast to partons, photons do not carry color 
charge and hence do not interact strongly when traversing the 
medium~\cite{ppg042}. The distribution of hard scattering vertices 
sampled by direct photon-triggered correlations is thus unbiased by the 
trigger condition.  Suppression of the opposite jet is averaged 
over all path-lengths given by the distribution of hard scattering 
vertices. Second, at the Born level, direct photon production in 
$p$+$p$ and A+A collisions is dominated by the QCD Compton scattering 
process, $q$+$g\rightarrow q$+$\gamma$, and the photon momentum in the 
center-of-mass frame is exactly balanced by that of the recoil 
quark.  Although higher order effects and other complications to 
this idealized picture such as Next-to-Leading Order (NLO) 
$2\rightarrow 3$ ``fragmentation" photons or soft gluon radiation 
must be considered, the level of suppression can then be related 
directly to the energy loss of a parton of known initial momentum. 
In this way, the average path-length of the away-side parton may 
then be varied in a well controlled manner by selecting events of 
various momentum differences between the $\gamma$-$h$ pair.

For this reason, the $\gamma$+jet channel has long been considered 
the ``golden channel'' for studying parton energy loss 
\cite{Wang:1996pe,Wang:1996yh}.  Neglecting the above mentioned 
complications, specifically effects like transverse momentum 
broadening (the $k_{\rm T}$ effect) and parton-to-photon 
fragmentation, back-to-back $\gamma$-$h$ 
correlations in elementary collisions directly measure the 
fragmentation function of the recoil jet since $z \equiv 
p^h/p^{\rm jet} \approx p^h/p^{\gamma}$. In the standard picture of 
energy loss, partons are likely to lose some fraction of their 
energy in the medium, but are likely to fragment outside the 
medium.  Hence, the parton energy loss can be considered an 
effective modification to the fragmentation function.  Such a 
picture may be tested using $\gamma$-$h$ correlations in nuclear 
collisions. Complementary baseline measurements in $p$+$p$ 
collisions are used to test the theoretical description of 
correlations in vacuum and to constrain possible contributions from 
higher order processes.  Comprehensive reviews of direct photon 
phenomenology and data from elementary collisions may be found in 
\cite{owensrev, ferbel, Vogelsang:1997cq}.

\section{Detector Description and Particle Identification}

The data were taken with the PHENIX detector 
\cite{overview} using approximately 950 million Au+Au minimum bias 
events from the 2004 data set and 471 million photon-triggered 
events from the 2005 and 2006 $p$+$p$ data sets corresponding to 
integrated luminosities of 3 (2005) and 10.7 (2006) pb$^{-1}$. 
The Beam-Beam Counters (BBC)~\cite{global}, which are used to 
trigger the minimum bias data, select 92\% of the total inelastic 
cross section.  
In Au+Au the BBC and 
Zero-Degree Calorimeters (ZDC) were used for offline minimum bias 
event selection and centrality determination.  In $p$+$p$ 
collisions a high energy photon trigger, defined by coincidence 
between the BBC and a high energy Electromagnetic Calorimeter (EMCal) 
tower hit, was utilized. This EMCal based trigger \cite{Adler:2003pb} 
had an efficiency of $>$ 90\% for events with photons and \piz\ 
with energies in the range used in the analysis and within the 
detector's geometric acceptance.

The PHENIX central arms, each covering $\pm 0.35$ units of 
pseudorapidity around midrapidity and $90^\circ$ in azimuth, 
contain charged-particle tracking chambers and electromagnetic 
calorimeters~\cite{centralarm}.  The EMCal \cite{emcal} consists of 
two types of detectors, six sectors of lead-scintillator (PbSc) 
sampling calorimeters and two of lead-glass (PbGl) \v{C}erenkov 
calorimeters measuring EM energy with intrinsic resolution 
$\sigma_E/E = 8.1\%/\sqrt(E) \oplus 2.1\%$ and $5.9\%/\sqrt(E) 
\oplus$ 0.8\% respectively.  The fine segmentation of the EMCal 
($\Delta\eta \times \Delta\phi \sim 0.01 \times 0.01$ for PbSc and 
$\sim 0.008 \times 0.008$ for PbGl) allows for the reconstruction 
of $\pi^0$'s and $\eta$'s in the $2 \gamma$ decay channel out to 
\pt\ of 20 GeV/c. The details of direct photon, \piz\, and $\eta$ 
meson detection and reconstruction within PHENIX have been 
described previously \cite{ppg042, ppg054, ppg055}.  Photon 
candidates with very high purity ($>$ 98\% for energies $> 5$ GeV) 
are selected from EMCal clusters with the use of cluster shower shape 
and charged particle veto cuts.  Two-photon \piz\ and $\eta$ 
candidates are selected from photon pairs with pair invariant mass 
in the appropriate \piz\ or $\eta$ mass range.  Combinatorial 
$2\gamma$ background is reduced with cuts on energy asymmetry 
$\alpha_{12} = |E_1 - E_2|/ (E_1 + E_2)$, described in detail 
below.  Some fraction of \piz\ with \pt\ starting at $\approx$ 13 
GeV/c (in the PbSc detector) will appear as a single merged 
cluster, but with anomalous shower shape, and thus are removed from 
the analysis.  The \piz and $\eta$ mesons in the \pt\ range from about 4 
to 17 GeV/c and photons between 5 and 15 GeV/c are used in this 
analysis.  For $\gamma$ \pt\ between 13--15 GeV/c there is a $< 2$\% 
contribution of merged \piz\ cluster contamination, however this 
together with all sources of non-photon contamination are found to 
have a negligible impact on the two-particle correlation analysis 
of this report.  Direct photons and their two-particle correlations 
are obtained by statistical subtraction of the estimated meson 
(mainly \piz) decay photon contribution from the inclusive photon 
and $\gamma$-$h$ samples.

Charged hadrons are detected with the PHENIX tracking system 
\cite{tracking} which employs a drift chamber in each arm spanning 
a radial distance of 2.0--2.4 m from the beam axis with a set of 
pixel pad chambers (PC1) directly behind them.  The momentum 
resolution was determined to be $\delta p/p = 0.7\% \oplus 1.0 \%p$ 
where $p$ is measured in GeV/c. Secondary tracks from decays and 
conversions are suppressed by matching tracks to hits in a second 
pad chamber (PC3) at distance of $\sim 5.0$ m. Track projections to 
the EMCal plane are used to veto photon candidates resulting from 
charged hadrons that shower in the EMCal.

\section{Method}

\subsection{Two-Particle Correlations}

Two-particle correlations are constructed by measuring the yield of 
particle pairs as a function of the measured azimuthal angle 
between photon or parent meson triggers and charged hadron 
partners. The correlation function, $C(\Delta\phi) \equiv 
N_{\rm real}^{\rm pair}(\Delta\phi)/N_{\rm mixed}^{\rm pair}(\Delta\phi)$, corrects 
for the limited acceptance of $\gamma$-$h$ or meson-hadron pairs 
by dividing the distribution in real events $N_{\rm real}^{\rm pair}$ by 
the mixed event distribution $N_{\rm mixed}^{\rm pair}$.  The correlation 
function is decomposed utilizing a two-source model of pair yields 
coming from two-particle jet correlations superimposed on a 
combinatorial background yield from an underlying event. The 
underlying event in Au+Au is known to have an azimuthal asymmetry 
of harmonic shape quantified in the elliptic flow parameter $v_2$ 
\cite{ppg098,ppg046}. This flow represents a harmonic modulation of 
the \dphi\ distribution of this underlying event, such that the 
flow-subtracted jet correlation signal is encoded in the jet pair 
ratio function, $JPR(\Delta\phi) \equiv C(\Delta\phi) - \xi 
(1+2\langle v_2^{\gamma}\rangle \langle v_2^h \rangle 
\cos{2\Delta\phi})$, using the notation of \cite{ppg083}, where 
$\langle v_2 \rangle$ is the average single-particle $v_2$.

Two methods of determining the background level $\xi$, known as 
Zero-Yield at Minimum (ZYAM) and Absolute Normalization (ABS) 
respectively were applied to the Au+Au data. Both methods are 
described in detail in previous PHENIX publications ~\cite{ppg083}, 
see also~\cite{ppg033,ppg083,McCumber:2005cv} (ABS) and 
~\cite{ppg032} (ZYAM). ZYAM assigns the level of zero jet yield and 
hence $\xi$ to the minimum point of the correlation function 
$C(\Delta\phi)$.  The ABS method uses the mean multiplicity of 
trigger-associated pairs in mixed events and a correction for 
finite centrality resolution to determine $\xi$.  Where ZYAM 
statistical precision is reasonable, the direct $\gamma$-h 
extraction of the two methods agree to within much better than the 
total uncertainties, typically within $\leq 20$\%.  The ABS method 
is chosen for the Au+Au results presented, as this method resulted 
in a more precise extraction of direct photon-jet pair yields at 
high trigger \pt\ where lack of statistics near $\Delta\phi = 
\pi/2$ severely impairs the ZYAM determination.  In the 
comparatively low multiplicity $p$+$p$ collisions, the underlying 
event originates from different physical mechanisms than in Au+Au 
and is known not to be well described by event-mixing.  Instead the 
correlation functions are normalized by fitting to a double 
Gaussian + constant function, corresponding to the ZYAM method 
\cite{ppg083}.

The results presented here are corrected for the associated charged 
hadron efficiency $\epsilon_h$ such that the quoted yields 
correspond to a detector with full azimuthal acceptance and $|\eta| 
< 0.35$ coverage.  No correction is applied for the $\Delta\eta$ 
acceptance of pairs.  Final results are presented in terms of the 
yield $Y$ of jet pairs per trigger, 
$Y \equiv A~JPR(\Delta\phi)/N_{\rm trigger}$ with the constant
$A = \int N^{\rm pair}_{\rm mixed}(\Delta\phi)/(2\pi \epsilon_h)$.

The magnitudes of elliptic flow were determined by measuring the 
distributions of inclusive photons, neutral pions, and charged 
hadrons as a function of the angle relative to the reaction plane, 
which was determined with the BBC's as described in \cite{ppg022}.
The \vtwo\ values measured for this 
analysis are consistent with previous PHENIX analyses 
\cite{ppg098,ppg046,ppg092}.  At high-\pt\ ( $\geq 6$ GeV/c) the 
measured \piz\ \vtwo\ values used in the determination of the decay 
photon \vtwo\ are fit to a constant function in order to reduce the 
effects of large statistical fluctuations.  The \pt\ independence 
of \vtwo\ of \piz's is motivated by recent preliminary data 
~\cite{kentaro_prelim} and also by the observed \pt\ independence 
of the \raa, since parton energy loss is expected to be the 
dominant mechanism for $v_2$ generation at 
high-\pt~\cite{Eskola:2004cr}.  It is also consistent with the 
findings of \cite{ppg092} which is direct measurement of \piz\ 
$v_2$ for the same dataset and is being published concurrently with 
this measurement.  Since, as discussed in that publication, the 
high-\pt\ functional behavior for this dataset cannot be 
well-constrained, the level of uncertainty we assign to the 
constant fit assumption increases with \pt.  

Table \ref{tab:v2vals} lists the $v_2$ values for the inclusive and 
\piz\ decay photons for all \pt\ ranges used, either the measurements, 
or for the highest \pt\ decay $v_2$ values from the constant fit value.  
For the fit values the fit errors are listed as statistical 
error, despite the inherent systematic correlation of the fit value 
across the \pt\ bins.  The decay photon $v_2$ is derived from the 
measured \piz\ $v_2$ by the same $p_{\rm T}^{\pi^0} \rightarrow p_{\rm 
T}^{decay \,\gamma}$ mapping procedure applied to the yields, described 
below.  It is assumed that the $v_2$ for other mesons which contribute 
decay photons ({\it e.g.} $\eta$) are the same as that of the \piz\ at 
high-\pt.  This assumption is well motivated for the \pt\ range 
considered ( $\gtrsim$ 4.5 GeV/c) under the expectation that the source 
of the high \pt\ azimuthal asymmetry $v_2$ is jet quenching-induced 
suppression, already measured to be the same for a variety of mesons 
($e.g.$ $\eta$\ itself \cite{ppg055}) and by data measurements for other 
high \pt\ $v_2$ which confirm the expectation \cite{Huang:2008vd} for 
other hadrons.

\begingroup \squeezetable
\begin{table}[bth]
\caption{\label{tab:v2vals}
$v_2$ values used in the jet function extraction for
inclusive and decay photons in Au+Au collisions.} 
\begin{ruledtabular} \begin{tabular}{cccccccccc}
& &&\multicolumn{3}{c}{Inclusive $\gamma$} && \multicolumn{3}{c}{Decay $\gamma$} \\
Centrality &  $p_{\rm T}^{\gamma}$ && \vtwo\ & Stat. & Sys. && \vtwo\ & Stat. & Sys. \\
\hline
& 5--7   && 0.053 & $\pm$0.009 & $\pm$0.011 && 0.084 & $\pm$0.009 & $\pm$0.004\\
& 7--9   && 0.047 & $\pm$0.022 & $\pm$0.015 && 0.069 & $\pm$0.018 & $\pm$0.003\\
0--20\% 
& 9--12  && 0.024 & $\pm$0.042 & $\pm$0.017 && 0.069 & $\pm$0.020 & $\pm$0.003\\
& 12--15 && 0.064 & $\pm$0.096 & $\pm$0.094 && 0.069 & $\pm$0.023 & $\pm$0.003\\
\hline
& 5--7   && 0.096 & $\pm$0.010 & $\pm$0.005 && 0.155 & $\pm$0.011 & $\pm$0.036\\
& 7--9   && 0.079 & $\pm$0.027 & $\pm$0.011 && 0.105 & $\pm$0.019 & $\pm$0.025\\
20--40\% 
& 9--12  && 0.025 & $\pm$0.050 & $\pm$0.049 && 0.105 & $\pm$0.020 & $\pm$0.025\\
& 12--15 && 0.287 & $\pm$0.128 & $\pm$0.104 && 0.105 & $\pm$0.023 & $\pm$0.024\\
\hline
& 5--7   &&  0.143 & $\pm$0.023 & $\pm$0.035 && 0.136 & $\pm$0.022 & $\pm$0.010\\
& 7--9   &&  0.146 & $\pm$0.064 & $\pm$0.026 && 0.126 & $\pm$0.039 & $\pm$0.008\\
40--60\%
& 9--12  &&  0.162 & $\pm$0.126 & $\pm$0.252 && 0.126 & $\pm$0.042 & $\pm$0.008\\
& 12--15 && -0.603 & $\pm$0.308 & $\pm$0.191 && 0.126 & $\pm$0.046 & $\pm$0.008\\
\end{tabular} \end{ruledtabular} 
\end{table} \endgroup

\clearpage

\subsection{ Direct $\gamma$-Hadron Correlation Subtraction}

A direct photon is defined here to be any photon not from a decay 
process.  Direct photons cannot be identified in Au+Au with 
reasonable purity on an event-by-event basis due to the large 
background of meson decay in the \pt\ range of the analysis and the 
inability to use isolation cuts in the high multiplicity Au+Au 
environment.  Thus both direct $\gamma$ and $\gamma$-$h$ pairs must 
be determined from the already mentioned statistical subtraction 
procedure, which is therefore consistently used in this report for 
both the $p$+$p$ and Au+Au.

Single direct photons have previously been measured in PHENIX, for 
Au+Au ~\cite{ppg042}, and $p$+$p$ \cite{ppg060}.  In these 
analyses, the estimated yield of decay photons $N_{\rm decay}^{\gamma}$ 
is subtracted from a measured sample of inclusive photons 
$N_{\rm inclusive}^{\gamma}$ resulting in the direct photon yield.  
These measurements serve as an input to the current analysis, as 
they fix the fraction of the photon triggers which are expected to 
be direct.  This fraction is quantified by the 
fraction \rg $\equiv N_{\rm inclusive}^{\gamma}/N_{\rm decay}^{\gamma}$. 
The \rg\ values used in this analysis are extracted from previous 
PHENIX measurements, \cite{Isobe:2006bf,Isobe:2007ku} by 
interpolating to obtain the \pt\ binning used in this analysis.  
These interpolated values together with the error estimations are 
tabulated in Table \ref{tab:rgamma}.

\begin{table}[tbh]
\caption{\label{tab:rgamma}
Extracted \rg\ values used as input to direct 
$\gamma$-h per-trigger yield subtraction (Equation 
\ref{eq:subtraction}).  
These values are interpolated from previous PHENIX measurements 
as described in the text. 
} 
\begin{ruledtabular} \begin{tabular}{ccccc}
Centrality & $p_{\rm T}^{\gamma}$ &  $R_{\gamma}$  & Stat. & Sys.\\
\hline
 & 5-7 & 1.77 & $\pm$0.09 & $\pm$0.06 \\
 & 7-9 & 2.45 & $\pm$0.09 & $\pm$0.18 \\
0-20\% 
 & 9-12 & 2.99 & $\pm$0.11 & $\pm$0.41 \\
 & 12-15 & 3.66 & $\pm$0.24 & $\pm$0.68 \\
\hline
 & 5-7 & 1.46 & $\pm$0.10 & $\pm$0.04 \\
 & 7-9 & 1.85 & $\pm$0.10 & $\pm$0.12 \\
20-40\% 
 & 9-12 & 2.30 & $\pm$0.12 & $\pm$0.28 \\
 & 12-15 & 2.35 & $\pm$0.20 & $\pm$0.44 \\
\hline
 & 5-7 & 1.30 & $\pm$0.09 & $\pm$0.05 \\
 & 7-9 & 1.52 & $\pm$0.07 & $\pm$0.13 \\
40-60\% 
 & 9-12 & 1.85 & $\pm$0.10 & $\pm$0.30 \\
 & 12-15 & 1.94 & $\pm$0.24 & $\pm$0.36 \\
\hline
  & 5-7 &1.18 & $\pm$0.01 & $\pm$0.06 \\
  & 7-9 &1.33 & $\pm$0.01 & $\pm$0.05 \\
 $p$+$p$ 
  & 9-12 &1.53 & $\pm$0.03 & $\pm$0.05 \\
  & 12-15 &1.79 & $\pm$0.09 & $\pm$0.07 \\
\end{tabular} \end{ruledtabular} \end{table}

The per-trigger yield of inclusive $\gamma$-$h$ pairs $Y_{\rm inclusive}$ is
simply the weighted average of the contributions from decay and direct
photon triggers,
\begin{equation}
Y_{\rm inclusive} = \frac{N_{\rm direct}^\gamma Y_{\rm direct} +  N_{\rm decay}^ \gamma
Y_{\rm decay}}{N_{\rm inclusive}^\gamma}.
\end{equation}
\noindent Having already determined \rg, $Y_{\rm direct}$ may then be
obtained by simple manipulation of the above terms resulting in
statistical subtraction involving only per-trigger yields as
follows. The decay photon per-trigger yield is subtracted from that
of inclusive photons according to:
\begin{equation}
  Y_{\rm direct} = \frac{R_{\gamma}  Y_{\rm inclusive} - Y_{\rm decay}}{R_{\gamma} -1}
\label{eq:subtraction}
\end{equation}
The direct $\gamma$ or direct $\gamma$-h pair yields do not, by
definition, exclude photons from jet fragmentation or medium induced
photon production.

\subsection{Extraction of Decay Photon Correlations}

The decay photon associated yields are estimated from the measured
\piz-h and $\eta$-h correlations through a calculation which
determines the decay correlations statistically from a Monte Carlo
(MC) based, pair-by-pair weighting procedure.   In this procedure
the decay $\gamma$-h pair yield
$N_{\rm decay}^{\gamma-h}(p_{\rm T}^{\gamma})$ is constructed 
by a weighted integral over all \piz-h and $\eta$-h pairs.  In what
follows, we will first describe the procedure schematically,
describing the ingredients and how they are obtained.  We then give
a more exact description and associated formula representing exactly
how the weighting was performed in the measurement.  Schematically
the procedure may be expressed as a convolution of several factors
according to the following relation, wherein for simplicity we only
consider photons from \piz\ decay, although the procedure is also
applied to $\eta$ decay photons.
\begin{equation}
N_{\rm decay}^{\gamma-h}(p_{\rm T}^{\gamma}) =  \int
\frac{\epsilon_{\gamma}(p_{\rm T}^{\gamma},p_{\rm T}^{\pi}) \otimes \mathcal{P}(p_{\rm T}^{\gamma},p_{\rm T}^{\pi})}{\epsilon_{\pi}(p_{\rm T}^{\pi})} \otimes N^{\pi-h}(p_{\rm T}^{\pi}) \label{eq:decscheme}
\end{equation}
\noindent where $\epsilon_{\pi}$ and $\epsilon_{\gamma}$ are the \piz\ and 
single $decay$ photon efficiencies, respectively, and $\mathcal{P}$ is the 
decay probability density, each of which is addressed in turn below.

\begin{figure}[tbh]
\includegraphics[width=1.0\linewidth]{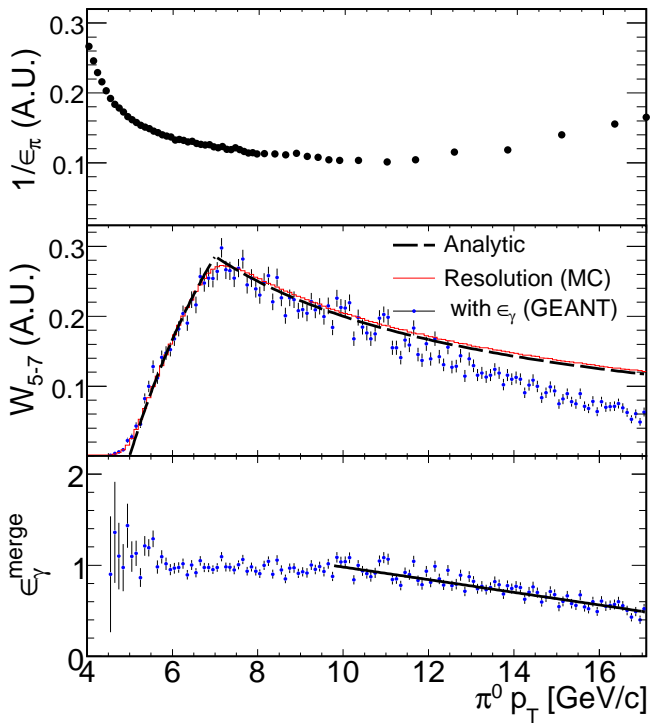}
\caption{\label{fig:shark} The weight factors used to obtain decay
correlations from parent meson correlations.  Top: \piz\
reconstruction efficiency correction, 1/$\epsilon_{\pi}$.  Middle:
Decay probability function, $W_{ab}$, for 5--7 GeV/c decay photons
from \piz\ derived analytically (black line), using the detector
acceptance and resolution smearing (red line) and including the
single decay photon efficiency, $\epsilon_{\gamma}$ from a {\sc GEANT}
simulation (blue points). Bottom: $\epsilon_{\gamma}^{\rm merge}$
obtained by taking ratio of the blue points to red curve in the
previous panel. }
\end{figure}

First, since the starting point is the uncorrected raw meson-h pair 
yield $N^{\pi-h}$, a correction for the parent meson reconstruction 
efficiency, $\epsilon_{\pi}(p_{\rm T}^{\pi})$, is applied to the raw 
\piz's as a function of \pt\ in order to account for the \piz\ daughter 
photons in the inclusive sample whose sisters lie outside the PHENIX 
acceptance or are otherwise undetected.  Both efficiencies 
$\epsilon_{\gamma}$, and $\epsilon_{\pi}$ in Equation \ref{eq:decscheme} 
are also evaluated as a function of the position in the calorimeter 
along the beam direction, however this dependence mostly cancels in the 
ratio $\epsilon_{\gamma}/\epsilon_{\pi}$ and therefore is suppressed for 
clarity. $\epsilon_{\pi}(p_{\rm T}^{\pi})$ is determined by dividing the 
raw number of \piz's $N^\pi(p_{\rm T}^{\pi})$ obtained in the same data 
sample by the PHENIX published \piz\ invariant yields 
~\cite{ppg063,ppg055,ppg080} assuming no pseudorapidity dependence over 
the narrow PHENIX acceptance.  The top panel in Fig.~\ref{fig:shark} 
illustrates, for the example of central Au+Au events, the \piz\ 
efficiency correction factor 1/$\epsilon_{\pi}(p_{\rm T}^{\pi})$.  The 
correction rises at small \pt\ due to a \pt-dependent pair energy 
asymmetry cut designed to reduce combinatorial 2$\gamma$ pairs 
reconstructed as real \piz's.  This cut, along with the effects of any 
remaining background, is described below.  At large \pt\, 
1/$\epsilon_\pi(p_{\rm T}^{\pi})$ rises again due to losses from cluster 
merging.

Second, the effect of decay kinematics is evaluated by determining the 
probability density, $\mathcal{P}(p_{\rm T}^{\gamma},p_{\rm T}^{\pi})$, 
for the decay of a \pt-independent distribution of \piz's.  
$\mathcal{P}(p_{\rm T}^{\gamma},p_{\rm T}^{\pi})$ represents the 
relative probability of a \piz\ of \pt=$p_{\rm T}^{\pi}$, to decay into 
a photon of $p_{\rm T}^{\gamma}$.  For a perfect detector, this function 
is calculable analytically.  A simple fast MC generator implements the 
PHENIX acceptance and uses Gaussian smearing functions to simulate 
detector resolution according to the known EMCal energy and position 
resolution. Occupancy effects give rise to an additional smearing of the 
\piz\ and $\eta$ invariant masses.  This effect is included in the MC by 
tuning the resolution parameters to match the \piz\ peak widths observed 
in data.  False reconstruction of \piz's and $\eta$'s from combinatorial 
matches are either subtracted or assigned to the systematic 
uncertainties as discussed below.

Finally, we wish to estimate the decay photon contribution to the 
measured raw inclusive photon sample which differs from the true decay 
photon distribution by the single decay photon efficiency, 
$\epsilon_{\gamma}(p_{\rm T}^{\pi})$.  At intermediate \pt, 
$\epsilon_{\gamma}(p_{\rm T}^{\pi})$ depends only on the photon momentum 
and is included already implicitly by the fast MC simulation described 
above to produce $\mathcal{P}(p_{\rm T}^{\gamma},p_{\rm T}^{\pi})$. 
Thus, it is useful to think of them as a single factor $W(p_{\rm 
T}^{\gamma}, p_{\rm T}^{\pi}) \equiv \mathcal{P}(p_{\rm T}^{\gamma}, 
p_{\rm T}^{\pi}) \epsilon_{\gamma}(p_{\rm T}^{\gamma}, p_{\rm T}^{\pi})$ 
At high-\pt, on the other hand, an efficiency loss is incurred by 
photons from \piz's whose showers merge into a single cluster in the 
calorimeter and are rejected by the shower-shape cut.  As a consequence, 
the fraction of photons that are direct is artificially enhanced in the 
sample of reconstructed photon clusters.  The single decay photon 
efficiency depends on both the parent and daughter \pt\ and is evaluated 
in a {\sc GEANT} simulation.  In principle the convolution of both 
$\mathcal{P}(p_{\rm T}^{\gamma},p_{\rm T}^{\pi})$ and 
$\epsilon_{\gamma}(p_{\rm T}^{\gamma},p_{\rm T}^{\pi})$, $W(p_{\rm 
T}^{\gamma}, p_{\rm T}^{\pi})$, could be extracted as one function from 
the {\sc GEANT} simulation, but obtaining large enough MC statistics necessary 
to properly parameterize the above mentioned EMCal z position dependence 
of the $\epsilon_{\pi,\gamma}$ corrections is only feasible with the 
fast MC.  Thus only the efficiency loss by cluster merging for photons 
$\epsilon_{\gamma}^{\rm merge}$ is taken from the {\sc GEANT}.  The bottom panel 
of Fig.~\ref{fig:shark} shows $\epsilon_{\gamma}^{\rm merge}(p_{\rm 
T}^{\pi})$ evaluated from the {\sc GEANT} simulation .

Since we wish to construct per-trigger yields, the same procedure 
described in Equation \ref{eq:decscheme} can be applied to find the 
estimated single decay photon trigger yield from the measured single 
\piz\'s, $i.e.$ replacing $N_{\rm decay}^{\gamma-h}$ with 
$N_{\rm decay}^{\gamma}$ and $N^{\pi-h}$ with $N^{\pi}$.  The exact 
application of schematic Equation \ref{eq:decscheme} then takes the form 
of a sum over all \piz-h pairs and single \piz\'s found in the data.  
Each \piz\ or \piz-h pair is given a weight which depends on \piz\ \pt.  
Operationally we now split this weight into two parts: 
$\epsilon_{\pi}(p_{\rm T}^\pi)$ discussed above and a factor 
$W_{ab}(p_{\rm T}^\pi)$.  The factor $W_{ab}$ is simply the end result 
of the fast MC-{\sc GEANT} combined calculation, the convolution of 
$\mathcal{P}$ and $\epsilon_{\gamma}$, including 
$\epsilon_{\gamma}^{\rm merge}$, averaged over a chosen decay photon bin of 
the range $a<p_{\rm T}<b$.  Thus in terms of the product $W(p_{\rm 
T}^{\gamma}, p_{\rm T}^{\pi})$ then $W_{ab}(p_{\rm T}^{\pi})$ is given 
by
\begin{equation}
W_{ab}(p_{\rm T}^{\pi}) = \int_a^b dp_{\rm T}^{\gamma} W(p_{\rm T}^{\pi},p_{\rm T}^{\gamma})
\end{equation}
\noindent Functions $W_{ab}(p_{\rm T}^{\pi})$ are defined for the four 
photon \pt\ bins used in the analysis, [a,b] = [5,7], [7,9], [9,12] and 
[12,15] GeV/c.  An example of $W_{ab}(p_{\rm T}^{\pi})$ for the 5-7 
GeV/c bin is shown in Fig.~\ref{fig:shark}.  Procedurally, we 
construct $W_{ab}$ as product of the fast MC curve shown in the middle 
panel and the linear fit discussed above to the bottom panel, 
$\epsilon^{\rm merge}_{\gamma}(p_{\rm T}^{\pi})$. Although a decay of \ptp\ 
$ < a $, the lower limit of the decay \pt\ bin, is kinematically 
disallowed, $W_{ab}$ is non-zero below this boundary when resolution 
effects are considered. For \ptp\ $> b$, $W_{ab}$ decreases as $\sim 
1/$\ptp, slowly enough that \piz's at values of \pt\ beyond the 
statistical reach of the data set contribute to the relevant decay 
photon \pt\ selections at a non-negligible rate. The \piz\ sample is 
truncated at $p_{\rm T} = 17$ GeV/c and extrapolated using power-law 
fits to the single and conditional \piz\ spectra to estimate a 
correction.  In the latter case, each associated hadron \pt\ range is 
fit independently.  The truncation avoids the high-\pt\ region where 
cluster merging effects are dominant and the $1/\epsilon^{\pi}$ 
correction factor becomes large.  Although the truncation corrections 
for the number of decay photons and decay $\gamma$-$h$ pairs are 
non-negligible, they mostly cancel in the per-trigger yield and are 
therefore typically $< 1 \% $, reaching a maximum value of 7\% for only 
the $12 <$ \ptg\ $ < 15 \otimes 3< $ \pth\ $< 5$ GeV/c bin.

With the weight functions $W_{ab}$ the entire set of \piz-hadron pairs 
and single \piz\ candidates (within a given range of \dphi, $\phi_1 < 
\Delta\phi < \phi_2$, defining each \dphi\ bin) are then summed over, 
once for each decay photon \pt\ bin, and the per-trigger yield is 
constructed for each of these decay \pt\ bins as

\begin{equation}
Y_{\rm decay}|_{a<p_{\rm T}^{\gamma}<b}^{\phi_1< \Delta\phi <\phi_2} =
 \frac{\sum\limits_{i=1-N^{\pi-h}}^{\phi_1< \Delta\phi_{\pi-h} < \phi_2}
 W_{ab}(p_{T i}^{\pi_i})/\epsilon_{\pi}(p_{\rm T}^{\pi_i})}
  {\sum\limits_{i=1-N^\pi}
 W_{ab}(p_{T i}^{\pi})/\epsilon_{\pi}(p_{\rm T}^{\pi_i})}
 \label{eq:sumdec}
\end{equation}

\noindent In this form it is clear that the normalization of the 
functions $\epsilon_{\pi}(p_{\rm T}^{\pi})$ and $W_{ab}(p_{\rm 
T}^{\pi})$ cancel out completely in the per-trigger yield, and therefore 
only their shapes versus $p_{\rm T}^{\pi}$ are important.  Hence in 
Fig.~\ref{fig:shark} the curves are shown with arbitrary units.  Also, 
as Equation \ref{eq:sumdec} implies, the angular deviation between the 
direction of a decay photon and its parent meson is ignored. The \dphi\ 
opening angle of a decay photon and hadron pair is taken to be the same 
as the $\Delta\phi_{\pi-h}$ of the parent \piz-h pairs. This 
approximation is tested in the fast MC and found to be extremely 
accurate since the distribution of angular deviation between a leading 
decay photon in a 2$\gamma$ decay and the parent mesons at these \piz\ 
momenta have an RMS around 0 of $\ll 0.01$ radians, and the smallest 
\dphi\ bins considered in the analysis are typically $\sim$ 0.1 radians 
or larger.

\subsection{\piz\ and $\eta$ Reconstruction}

\label{sec:pizetarec}

In $p$+$p$ collisions $Y_{\rm decay}$ is estimated using both
reconstructed \piz\ and $\eta$ mesons in invariant mass windows of
120--160  and 530--580 $\rm{MeV/c}^2$, respectively.  The
total decay per-trigger yield is calculated from

\begin{equation}
Y_{\rm decay} = (1-\delta^{\gamma}_{h/\pi^0}) Y_{\rm decay}^{\pi^0} +
\delta^{\gamma}_{h/\pi^0} Y_{\rm decay}^\eta \label{eq:etadec}
\end{equation}

\noindent where $\delta^{\gamma}_{h/\pi^0}$ is the ratio of the total 
number of decay photons to the number of decay photons from \piz.  
Based on the measurements of $\eta$ \cite{ppg055} and $\omega$ 
\cite{ppg064}, which together with the \piz\ account for $> 99 \%$ of 
decay photons, the value of $\delta^{\gamma}_{h/\pi^0}$ is determined to 
be $1.24 \pm 0.05$ in the high-\pt\ region covered by this analysis, 
independent of collision system and centrality. Note that the 
per-trigger yields for $\omega$ and other heavier meson triggers 
($\omega$,$\eta\prime$,$\phi$,...) are not measured and are taken to be 
equivalent to $Y_{\rm decay}^{\eta}$ in Equation \ref{eq:etadec}. This 
assumption was studied in PYTHIA and found to influence $Y_{\rm decay}$ at 
the level of $< 2\%$.  In Au+Au collisions correlations using $\eta$ 
triggers are not directly measured, but rather estimated from the 
$p$+$p$ measurement as discussed below.

\begin{figure}[tb]
\includegraphics[width=1.0\linewidth]{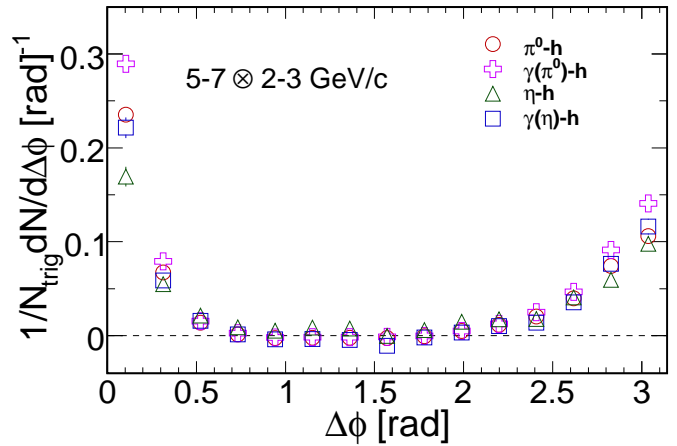}
\caption{\label{fig:JFdec} (color online) Examples of parent and
daughter per-trigger yields for the \piz\ and $\eta$ in $p$+$p$
collisions for \pt\ selection 5 $< $\ptg$ < $ 7 and 2 $< $\pth$ <$ 3
GeV/c. These correlation measurements are used to determine the total decay
photon per-trigger yield as described in the text.}
\end{figure}

Figure~\ref{fig:JFdec} shows the 
various components of the decay photon measurement in $p$+$p$.
In $p$+$p$ collisions the rate of combinatorial background photon pairs 
is reduced by only considering photons of \pt\ $> 1$ GeV/c resulting in 
background levels of $<$ 10\% for which no correction was applied.  The 
effect of such remaining pairs on $Y_{\rm decay}^{\pi^0}$ was evaluated to 
be negligible ($<$ 2\%) compared to the size of other uncertainties on 
the final $Y_{\rm direct}$ result using a detailed full PYTHIA test of the 
method which included \piz\ reconstruction with combinatorial photon 
pairs.  On the other hand, $\eta$ reconstruction has a much smaller 
signal-to-background of 1.4--1.6, depending on the \pt\ selection, even 
in the low multiplicity $p$+$p$ environment.  In this case, the 
per-trigger yield of the combinatorial photon pairs is estimated from 
photon pairs with invariant mass in ``sideband'' ranges of 400--460 and 
640--700 $\rm{MeV/c}^2$, beyond 3$\sigma$ of the $\eta$ peak.  The 
sideband contribution $Y_{\rm decay}^{\rm sideband}$ is then subtracted using 
the signal-to-background ratio $f_{\rm bkg}$ evaluated from gaussian + 
polynomial background fits to the invariant mass distributions according 
to $Y_{\rm decay}^{\rm signal} = Y_{\rm decay}^{\rm raw}/(1/f_{\rm bkg}+1) - 
Y_{\rm decay}^{\rm sideband}/f_{\rm bkg}$.  The yield $Y_{\rm decay}^{\rm sideband}$ is 
generated from the full meson to decay photon weighting function 
procedure (Equation \ref{eq:sumdec}).  The subtraction procedure was 
also tested in PYTHIA and the extracted and input per-trigger yields 
were found to agree to within 10\%.  

In Au+Au collisions the combinatorial rate for \piz\ reconstruction is 
substantially larger.  Correspondingly, a \pt\ dependent cut on the pair 
energy asymmetry $\alpha_{12} = |E_1 - E_2|/ (E_1 + E_2)$~\cite{ppg054}, 
visible in Fig.~\ref{fig:shark} with the smallest allowed asymmetry at 
the lowest \piz\ \pt\ values, is used to reduce this background.  With 
such cuts the signal-to-background in central events varies from 5:1 at 
its lowest, increasing to about 15:1 for the highest \pt\ selection.  
The effect of the combinatorial background is studied through 
examination of a similar sideband subtraction analysis as in the $p$+$p$ 
$\eta-h$ correlation extraction described, this time for \piz-h, using 
invariant mass ranges just outside the \piz\ peak region.  However no 
clear trend beyond non-negligible statistical limitations is observed, 
so no correction for the background is applied. Instead the maximum size 
of the effect (typically $\simeq$ 7\%) is included as source of 
systematic uncertainty on the decay yields and propagated to the final 
direct photon per-trigger yields.

In central Au+Au collisions the $\eta$ meson cannot be reconstructed 
with sufficient purity to measure its correlations.  Instead, a scaling 
argument is employed.  Motivated by the similar high-$p_{\rm T}$ 
suppression pattern shown by $\eta$ and \piz\ in Au+Au ~\cite{ppg055} 
and corresponding near equality of the $p$+$p$ and Au+Au $\eta/\pi^0$ 
ratios, the ratio $Y_{\gamma(\eta)}/Y_{\gamma(\pi^0)}$ is measured in 
$p$+$p$ and applied as a correction to the Au+Au $Y_{\gamma(\pi^0)}$. 
This is justified by the assumption that the jet fragmentation is 
primarily occurring outside the medium.  We do not attribute any 
additional uncertainty to this scaling beyond the 10\% sideband 
systematic and statistical uncertainties of the $\eta$ measurement in 
$p$+$p$.  However, to give an idea of the possible impact of this 
assumption, the total systematic error on $Y_{\rm decay}$ from all other 
sources would correspond to a variation of the Au+Au $Y_{\gamma(\eta)}$ 
by $\sim$ 50\%.  Given the similarity of the high-\pt\ suppression 
demonstrated by all light quark bound states measured thus far, this 
would correspond to a rather large change.

\section{Systematic Uncertainties}

There are four main classes of systematic uncertainty in the Au+Au data: 
elliptic flow, normalization of the underlying event (ABS), \rg, and the 
decay per-trigger yield estimate, the latter two of which are present in 
the $p$+$p$ data as well. Table \ref{tab:summary} lists the fractional 
contribution of each of these sources to the total systematic 
uncertainty on the direct photon per-trigger yields in the 20\% most 
central Au+Au and $p$+$p$ data. In the central Au+Au data the 
uncertainty at low $p_{\rm T}^{h}$ is dominated by the $v_2$ and 
correlation function normalization (ABS method) estimation due to large 
multiplicity of hadrons. At higher $p_{\rm T}^{h}$, but low trigger \pt, 
$p_{\rm T}^{t}$, the decay error dominates due to the two-photon 
combinatorial background for \piz\ reconstruction.  Finally, at large 
$p_{\rm T}^{h}$ and $p_{\rm T}^{t}$ the backgrounds responsible for both 
of these sources of uncertainty decrease and the uncertainty on 
$R_{\gamma}$, which is relatively constant, dominates.  In $p$+$p$ 
collisions the decay photon background forms a much larger fraction of 
the total photon sample. In this case, the decay uncertainty arises from 
the MC decay photon mapping procedure, the $\eta$ sideband subtraction 
and the $\eta/\pi^{0}$ ratio in approximately equal parts.  The yields 
associated with daughter photons are larger than for the meson parents 
because of feed-down from larger values of parent \pt, and hence, jet 
\pt.

The correction for single hadron efficiency $\epsilon_h(p_{\rm T}^h)$ 
varies as a function of collision system and centrality.  These 
corrections are obtained by finding the ratio of raw yields of hadrons 
obtained without the trigger condition in the same analysis ($i.e.$) 
with the same cuts as in the analysis, to the previous PHENIX published 
measurements of the corresponding charged hadron spectra. \cite{ppg023, 
ppg050}.  As in previous PHENIX two-particle correlation measurements, 
\cite{ppg032, ppg083}, this procedure has inherent uncertainties 
assigned as a \pt-independent 10\% uncertainty, on each system and/or 
centrality.

\begingroup \squeezetable
\begin{table}
\caption{\label{tab:summary}
Fractional contribution to the total systematic uncertainty for each of the main
sources of uncertainty in $p$+$p$ and 0-20\% Au+Au collisions.
}
 \begin{ruledtabular} \begin{tabular}{cccccccccc}
\ptg & \pth 
&& \multicolumn{4}{c}{Au+Au, Centrality 0-20 \%} & \multicolumn{2}{c}{$p$+$p$} \\
 (GeV) &  (GeV) && $R_{\gamma}$ & Decay & v2 & Norm.&& $R_{\gamma}$ & Decay  \\
\hline
 & 1-2 && 0.03 & 0.14 & 0.50 &0.33 && 0.14 & 0.86\\
5-7 
 & 2-3 && 0.02 & 0.32 & 0.46 & 0.20 && 0.21 & 0.79 \\
 & 3-5 && 0.02 & 0.71 & 0.18 & 0.10 && 0.05 & 0.95 \\
  \hline
 & 1-2 && 0.09 & 0.17 & 0.45 &0.29 && 0.22 & 0.78\\
7-9 
 & 2-3 && 0.10 & 0.35 & 0.38 & 0.17 && 0.25 & 0.75 \\
 & 3-5 && 0.09 & 0.61 & 0.18 & 0.13 && 0.21 & 0.79 \\
  \hline
 & 1-2 && 0.06 & 0.09 & 0.53 &0.33 && 0.19 & 0.81\\
9-12 
 & 2-3 && 0.26 & 0.25 & 0.33 & 0.16 && 0.30 & 0.70 \\
 & 3-5 && 0.46 & 0.30 & 0.13 & 0.10 && 0.35 & 0.65 \\
  \hline
 & 1-2 && 0.08 & 0.01 & 0.63 &0.29 && 0.21 & 0.79\\
12-15 
 & 2-3 && 0.21 & 0.14 & 0.48 & 0.17 && 0.02 & 0.98 \\
 & 3-5 && 0.22 & 0.14 & 0.39 & 0.25 && 0.10 & 0.90 \\
\end{tabular} \end{ruledtabular} \end{table}
\endgroup

\section{Results}

\subsection{Direct $\gamma$-h Per-Trigger Yields}

\begin{figure}[tb]
\includegraphics[width=1.0\linewidth]{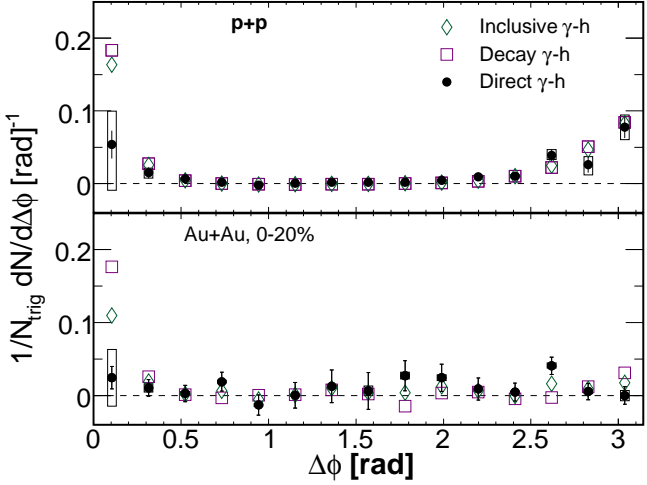}
\caption{\label{fig:JFdir} (color online) Examples of per-trigger
yields used in the direct photon correlation analysis for the 5
$< $\ptg$ < $ 7 and 3 $< $\pth$ <$ 5 GeV/c bin.  Top (bottom) panel:
Inclusive, decay and direct photon per-trigger yields in $p$+$p$
(0--20\% central Au+Au) collisions.}
\end{figure}

Figure \ref{fig:JFdir} shows examples of direct photon per-trigger 
yields in $p$+$p$ and central Au+Au collisions.  Also shown are the 
per-trigger yields for inclusive and decay photon triggers which are the 
ingredients in the statistical subtraction method as expressed in 
Equation \ref{eq:subtraction}. A clear away-side correlation is observed 
(\dphi\ $\simeq \pi$) for direct photons triggers in $p$+$p$. In Au+Au 
collisions the away-side correlation is suppressed for both decay and 
direct photon triggers. The near-side direct photon associated yields 
are small relative to that of decay photons, an expected signature of 
prompt photon production \cite{ferbel}.

\begin{figure}[tb]
\includegraphics[width=1.0\linewidth]{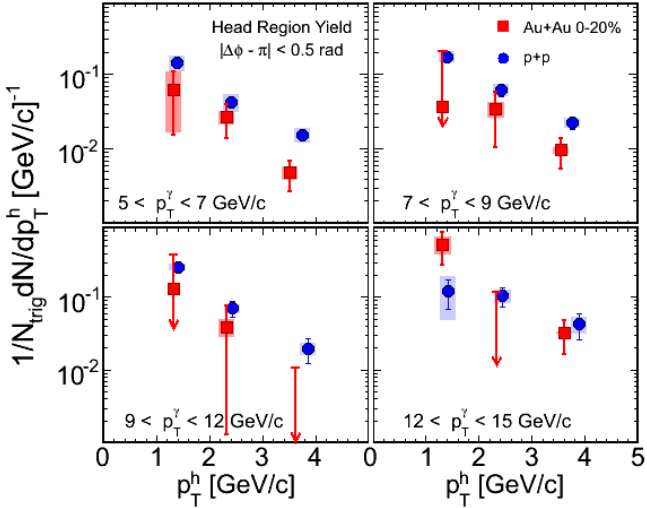}
\caption{\label{fig:fourpanelyields} (color online) Direct $\gamma$-$h$ 
per-trigger yields for the range $|\Delta\phi-\pi| < \pi/5$ radians 
\emph{vs.} associated hadron $p_{\rm T}$.  Four different direct 
$\gamma$ \pt\ ranges (indicated on the figure) are shown in the most 
central 20\% of Au+Au events and $p$+$p$ events. 
The upper limits are for 90\% confidence levels.
A \pt-independent uncertainty of 10\% due to the charged hadron 
efficiency correction is suppressed from the plot. }
\end{figure}

\begin{table}
\caption{\label{tab:yields}
Direct $\gamma$-$h$ per-trigger yields in 20\% most central 
Au+Au and in $p$+$p$ collisions.  An additional \pt-independent 
uncertainty of 10\% due to the charged hadron efficiency 
corrections is not shown. 
} 
\begin{ruledtabular} \begin{tabular}{cccccccc}
\ptg & \pth & 
$\langle z_{\rm T} \rangle$ & Yield & Stat & Sys & Total \\
 (GeV) & (GeV) & 
\multicolumn{5}{c}{Au+Au, Centrality 0--20\%} \\
\hline
 & 1-2  & 0.23  & 6.26e-02  & 4.72e-02  & 4.62e-02  & 6.60e-02 \\
5-7  & 2-3  & 0.41  & 2.68e-02  & 1.29e-02  & 5.68e-03  &1.41e-02 \\
  & 3-5  & 0.62  & 4.82e-03  & 2.13e-03  & 1.96e-03  &2.90e-03 \\
  \hline      
  & 1-2  & 0.17  & 3.71e-02  & 8.48e-02  & 5.59e-02  & 1.02e-01 \\
7-9  & 2-3  & 0.3  & 3.45e-02  & 2.39e-02  & 8.46e-03  &2.53e-02 \\
  & 3-5  & 0.46  & 9.63e-03  & 4.18e-03  & 1.96e-03  &4.62e-03 \\
  \hline      
  & 1-2  & 0.13  & 1.28e-01  & 1.34e-01  & 6.84e-02  & 1.51e-01 \\
9-12  & 2-3  & 0.23  & 3.94e-02  & 3.81e-02  & 1.01e-02  &3.94e-02 \\
  & 3-5  & 0.36  & -2.16e-03  & 6.29e-03  & 2.06e-03  &6.62e-03 \\
  \hline      
  & 1-2  & 0.1  & 5.31e-01  & 2.53e-01  & 1.49e-01  & 2.94e-01 \\
12-15  & 2-3  & 0.18  & -6.13e-03  & 6.99e-02  & 1.80e-02  &7.22e-02 \\
  & 3-5  & 0.28  & 3.25e-02  & 1.60e-02  & 2.52e-03  &1.62e-02 \\
\hline
  \multicolumn{7}{c}{$p$+$p$} \\
\hline      
 & 1-2  & 0.24  & 1.44e-01 & 9.93e-03 & 3.42e-02 & 3.56e-02 \\
5-7  & 2-3  & 0.43  & 4.22e-02 & 5.47e-03 & 1.20e-02 & 1.32e-02 \\
  & 3-5  & 0.66  & 1.55e-02 & 2.07e-03 & 3.26e-03 & 3.86e-03 \\
  \hline      
  & 1-2  & 0.18  & 1.73e-01 & 1.84e-02 & 2.88e-02 & 3.42e-02 \\
7-9  & 2-3  & 0.31  & 6.24e-02 & 1.11e-02 & 1.15e-02 & 1.60e-02 \\
  & 3-5  & 0.48  & 2.26e-02 & 4.53e-03 & 3.75e-03 & 5.88e-03 \\
  \hline      
  & 1-2  & 0.14  & 2.59e-01 & 2.99e-02 & 2.50e-02 & 3.90e-02 \\
9-12  & 2-3  & 0.24  & 7.01e-02 & 1.73e-02 & 1.00e-02 & 2.00e-02 \\
  & 3-5  & 0.38  & 1.94e-02 & 7.21e-03 & 3.77e-03 & 8.14e-03 \\
  \hline      
  & 1-2  & 0.11  & 1.20e-01 & 5.13e-02 & 7.22e-02 & 8.86e-02 \\
12-15  & 2-3  & 0.19  & 1.04e-01 & 3.11e-02 & 2.02e-02 & 3.71e-02 \\
  & 3-5  & 0.3  & 4.26e-02 & 1.62e-02 & 1.13e-02 & 1.97e-02 \\
\end{tabular} \end{ruledtabular} 
\end{table}

The away-side yields, integrated over $|\Delta\phi-\pi| < \pi/5$
radians, are shown in Fig.~\ref{fig:fourpanelyields} and Table
\ref{tab:yields} for $p$+$p$ and Au+Au collisions.  This range
roughly corresponds to the ``head region'' as defined
in~\cite{ppg083} and is chosen primarily to minimize the influence
of medium response which is thought to dominate the ``shoulder''
region further offset from $\Delta\phi$ = $\pi$. Additionally, the
acceptance and the signal itself are largest in this range so
statistical precision is maximized.  It should be noted that the
width of the jet correlation is larger than this interval.  We do
not make a correction for this effect, since we are primarily
concerned with the comparison of the yields from $p$+$p$ and Au+Au
collisions.  It should be noted, however, that in addition to parton
energy loss, any broadening of azimuthal correlations, whether by
hot or cold nuclear matter effects, will contribute to a suppression
in the yield in the head region.  Due to statistical and systematic
fluctuations, the subtraction of the decay-photon hadron pairs from
the inclusive $\gamma$-$h$ sample can result in a negative yield. In
this case 90\% confidence-level upper limits are given. In the case
that a positive yield is obtained, but the uncertainty is consistent
with 0, the lower bound of the error bar is also replaced with an
arrow. As noted in the figure caption, a 10\% \pt-independent
uncertainty due to the charged hadron efficiency corrections is not
shown.

\subsection{Suppression Factor $I_{\rm AA}$}

\begin{figure}[tb]
\includegraphics[width=1.0\linewidth]{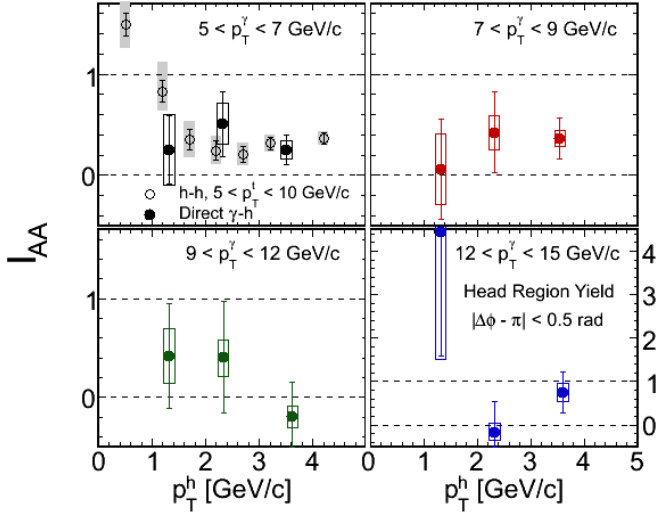}
\caption{\label{fig:fourpaneliaa} (color online) Ratio 
$I_{\rm AA}$ of the Au+Au to $p$+$p$ yields shown in 
Fig.~\protect\ref{fig:fourpanelyields}.  An additional 
\pt-independent uncertainty of 14\% due to the charged 
hadron efficiency corrections is not shown. 
}
\end{figure}

Departure from the vacuum QCD processes is quantified by 
$I_{\rm{\rm AA}}$, the ratio of \auau\ to \pp\ per-trigger yields:
\begin{equation} \label{eq:iaadef}
I_{\rm{\rm AA}}(p_{\rm T}^{\gamma},p_{\rm T}^{h}) =
\frac{Y^{Au+Au}(p_{\rm T}^{\gamma},p_{\rm T}^{h})}{Y^{p+p}(p_{\rm T}^{\gamma},p_{\rm T}^{h})}.
\end{equation}
\noindent Figure \ref{fig:fourpaneliaa} shows the \iaa\ values for 
all direct photon and associated hadron bins for the most central 
0--20\% of collisions.  The data points for which the subtraction 
resulted in a negative yield value (the 90\% confidence level upper 
limits) are included with standard 1-$\sigma$ uncertainties.  For 
the $p_{\rm T}^{\gamma}$ range 5--12 GeV/c, a significant 
suppression is observed in the $3<p_{\rm T}^h <5$ GeV/c bin in 
which the highest precision is obtained. At lower $p_{\rm T}^h$, 
where the background subtraction is largest, the data do not have 
the statistical precision to determine the degree to which the 
yields are suppressed.  $I_{\rm AA}$ for direct photon triggers is 
consistent to that of charged hadron triggers \cite{ppg083} as 
shown in the top left panel in which results with similar ranges of 
$p_{T,t}$ are compared.

\begin{figure}[tb]
\includegraphics[width=1.0\linewidth]{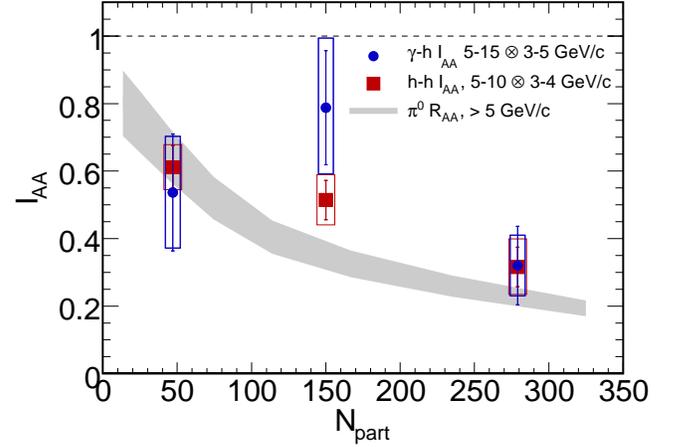}
\caption{\label{fig:integrals_vscent} (color online) $I_{\rm 
AA}$(\ptg)  integrated over the range 5 $<$ \ptg\ $<$ 15 GeV/c for 
associated hadrons of 3 $<$ \pth\ $<$ 5 GeV/c \emph{vs.} centrality 
compared to single \piz\ high-\pt\ $R_{\rm AA}$ (integrated over 
\pt\ $>$ 5 GeV/c)~\cite{ppg080}. An additional \pt-independent 
uncertainty of 14\% due to the charged hadron efficiency 
corrections is not shown.  }
\end{figure}

Figure~\ref{fig:integrals_vscent} shows the $I_{\rm AA}$ for the 
\pth\ = 3--5 GeV/c bin, integrated for all trigger \pt\ bins (\ptg\ 
= 5--15 GeV/c) and for three centrality bins, 0--20\%, 20--40\%, 
and 40--60\%.  For the most central bin, the suppression of the 
away-side direct photon per-trigger yield is clearly observed, 
$I_{\rm AA} = 0.32 \pm 0.12 ^{\rm stat} \pm 0.09 ^{\rm syst}$. 
Within large uncertainties we see that the $\gamma$-jet $I_{\rm 
AA}$ in this \pt\ range, dominated by moderate to high values of 
$z$ ($\equiv p^h/p^{\rm jet}$), is consistent with the single particle 
$R_{\rm AA}$ as a function of centrality, consistent with a 
scenario in which the geometry of suppression plays an important 
role as would be expected from a sample dominated by surface 
emission.

Figure~\ref{fig:integrals_vscent} also compares $I_{\rm AA}$ from a 
measurement of high-\pt\ dihadron ($h^\pm-h^\pm$) correlations 
\cite{ppg083} to the $\gamma$-jet result for similar $p_{T,t}$ 
selections.  The two results are remarkably similar in the most 
central bin.  This may indicate that surface emission is dominant 
for both samples in this $z$ region.  However it should be noted 
that the total uncertainties on either measurement are still quite 
large on a relative scale.  As explained in the introduction, the 
two measurements should be subject to different geometrical 
effects.  Disentangling such effects through precise comparisons of 
dihadron and $\gamma$-$h$ suppression should be pursued with future 
measurements with improved statistics.

\subsection{Towards the Fragmentation Function}

Using the distribution of charged hadrons opposite direct $\gamma$ 
triggers, parton energy loss may be studied directly as a departure 
from the (vacuum) fragmentation function.  In distinction to 
$\pi^0$-h correlations, where the away-side distribution is only 
sensitive to the integral of the fragmentation function (the 
average multiplicity of the away-side jet)~\cite{ppg029}, the 
away-side distribution for direct $\gamma$-h correlations provides 
a measurement of the full fragmentation function of the jet from 
the away-side parton.  To the extent that the transverse momentum 
of the away-side parton and the direct $\gamma$ are equal and 
opposite, as in leading order pQCD, the fragmentation function 
of the jet from the away-parton should be 
given to a good approximation by the $x_E$ distribution,
\begin{equation}
    \label{eq:xe}
x_E=\frac{-\vec{p}_{\rm T}^{\ t}\cdot\vec{p}_{\rm T}^{\ h}}{|\vec{p}_{\rm T}^{\
t}|^2}=
\frac{-p_{\rm T}^{h} \cos{\Delta\phi}}{p_{\rm T}^{t}} 
\end{equation}
\noindent where the transverse momentum of the trigger $p_{\rm 
T}^{t}$ = \ptg\ in the case of $\gamma$-h correlations. The reasons 
why the scaling variable $x_E$ is an approximation to, rather than 
exact measure of, the fragmentation variable of the away-side jet 
with momentum $z_a$ are: i) the away-side parton does not generally 
balance longitudinal momentum with the trigger $\gamma$, although 
it is restricted by the $\Delta \eta$ acceptance of the detector; 
ii) the transverse momenta of the $\gamma$ and away parton do not 
exactly balance. The transverse momentum imbalance was discovered 
at the CERN-ISR using $x_E$ distributions~\cite{DellaNegra} and 
originally attributed to an ``intrinsic'' transverse momentum 
$k_{\rm T}$ of each of the initial colliding 
partons~\cite{Feynman:1977yr}, but now understood to be due to 
``resummation'' of soft-gluon 
effects~\cite{Kulesza:2002tx,aurenche}.

\begin{figure}[tb]
\includegraphics[width=1.0\linewidth]{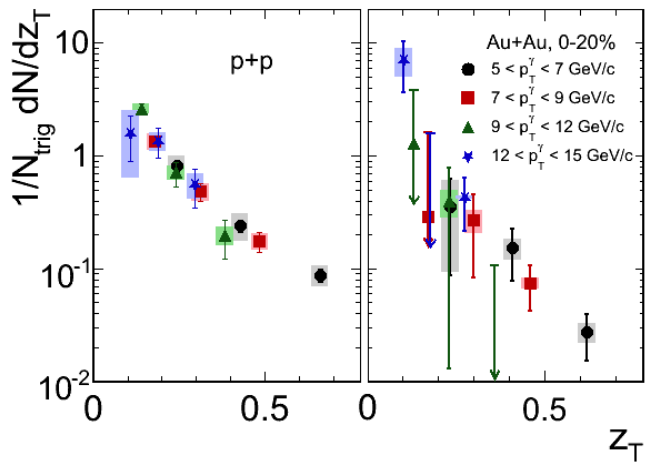}
\caption{\label{fig:xE}
$z_{\rm T}$ distributions $dN/dz_{\rm T}$ from the direct photon 
associated yields in $p$+$p$ (left) and 
0--20\% Au+Au (right) collisions. }
\end{figure}

The validity of the approximation $x_E\approx z_a$ can be tested by 
observing identical $x_E$ distributions for different values of 
trigger \ptg\ ($x_E$ scaling), in which case one would accept the 
$x_E$ distribution in $\gamma$-h correlations as the quark 
fragmentation function from the reaction $q+g\rightarrow q+\gamma$ 
without need of correction.  We approximate $x_E$ by $z_{\rm T}$, 
the ratio of the mean associated \pth\ to mean trigger \pt\ for 
each \ptg\ bin.\footnote{The reader is advised to carefully 
distinguish this variable $z_{\rm T} = \langle p_{\rm T}^{h} 
\rangle / \langle p_{\rm T}^{t} \rangle$ from our previous notation 
used in~\cite{ppg029} of $z_t = p_{\rm T}^{t}/\hat{p}$, which is 
the fraction of jet momentum $\hat{p}$ contained in the trigger 
particle.} The $\mean{p_{\rm T}^{\gamma}}$ for the four trigger 
bins are: 5.66, 7.75, 10.07, 13.07 GeV/c, close to the values 
obtained from a fit to the direct-$\gamma$ invariant cross section 
of the form $p_{\rm T}^{-6.5}$~\cite{ppg060}. 

\begin{figure}[tb]
\includegraphics[width=1.0\linewidth]{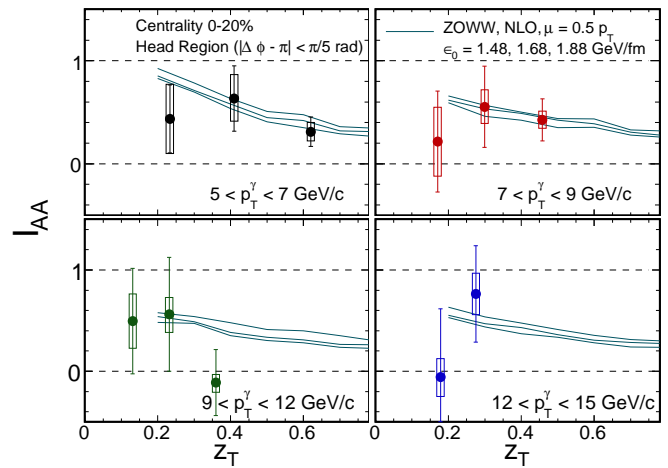}
\caption{\label{fig:iaawpred} (color online) $I_{\rm AA} (z_{\rm 
T})$ for the 20 \% most central Au+Au data compared to predictions 
from an energy loss calculation ~\cite{ZOWWpriv}. An additional 
\pt-independent uncertainty of 14\% due to the charged hadron 
efficiency corrections is not shown. }
\end{figure}

Figure \ref{fig:xE} shows the $z_{\rm T}$ distributions 
for $p$+$p$ and Au+Au collisions.
The $p$+$p$ data (Fig.~\ref{fig:xE}a) exhibit reasonable $z_{\rm 
T}$ scaling so that the measured distribution should represent the 
away-side jet fragmentation function. A fit of this data to a 
simple exponential $ (N e^{-b z_{\rm T}})$ gives an acceptable 
$\chi^2/dof=12.8/10$ with a value $b=6.9\pm0.8$ which is consistent 
with the quark fragmentation function, parameterized~\cite{ppg029} 
as a simple exponential with $b=8.2$ for $0.2<z< 1.0$, and 
inconsistent with the gluon fragmentation function value of 
$b=11.4$.  It should, however, be recalled that the data do not 
cover the full extent of the away peak, only 
$|\Delta\phi-\pi|<\pi/5$ radians, and that possible variations of 
the widths of the peaks in both the $p$+$p$ data and the Au+Au data 
with $p^{\gamma}_{\rm T}$ and $p^{h}_{\rm T}$ have not been taken 
into account in the present analysis.  Additionally a more detailed 
analysis, differential in trigger \pt, is necessary to study 
trigger \pt\ dependent effects which can influence the 
fragmentation function fit values \cite{ppg029}.

In central Au+Au collisions, the fragmentation function may be 
modified by the medium\footnote{See Equation 1 in \cite{ZOWW2}}, 
so that $z_{\rm T}$ scaling should not hold except in two special 
cases: i) pure surface emission or punch-through where the 
away-side jets are not modified---the $z_{\rm T}$ distribution will 
be suppressed, but will have the same shape as in $p$+$p$ 
collisions; ii) constant fractional energy loss of the away 
jet---the $z_{\rm T}$ scaling will be preserved in Au+Au collisions 
but with a steeper slope than in $p$+$p$ collisions. The Au+Au data 
(Fig.~\ref{fig:xE}b) are consistent with $z_{\rm T}$ scaling with 
the same shape as the $p$+$p$ data, a value of $b=5.6\pm 2.2$ and 
excellent $\chi^2/dof=10.1/10$ for the simple exponential fit. The 
point at lowest $z_{\rm T}=0.11$ for Au+Au is 1.6 standard 
deviations above the fit, suggesting that improved statistics will 
permit the observation of any non-surface emission.

\subsection{Model Comparison}

Several authors have reported predictions for $\gamma$-jet in heavy ion 
collisions \cite{galegjet,ZOWWpriv,arleoraa,renkgjet}.  As a 
demonstration of the how such calculations can be compared to the 
data, the \iaa\ values as a 
function of $z_{\rm T}$ are compared to energy loss predictions 
\cite{ZOWWpriv} in Fig.~\ref{fig:iaawpred}.  The calculation uses 
effective fragmentation functions to parameterize the energy loss 
in terms of a parameter $\varepsilon_{0}$ which is expected to be 
proportional to the initial gluon density~\cite{ZOWW2}.  The model 
calculates the energy-loss of the leading parton, and neglects the 
contribution the gluon radiation and medium response which may 
dominate at low values of $z$.  The data is well reproduced by the 
model over the range of values of $\varepsilon_{0}$ provided, 
1.48--1.88 GeV/fm.  This corresponds roughly to the range of 
$\varepsilon_{0}$ allowed by comparison to the PHENIX \piz\ $R_{\rm 
AA}$ data of $1.9 ^{+0.2}_{-0.5}$ \cite{ppg079}.

It should be noted that the calculation rejects fragmentation 
photons with an isolation cut. Such a procedure has not yet been 
demonstrated in central Au+Au data, although doing so would help to 
eliminate beyond-leading-order effects.

\section{Conclusions}

We have presented the first direct $\gamma$-h measurements in Au+Au 
and $p$+$p$ collisions at RHIC. A significant suppression of 
$I_{\rm AA} = 0.32 \pm 0.12 ^{\rm stat} \pm 0.09 ^{\rm syst}$ for 
the away-side charged hadron yield in the range $3 < p_{\rm T}^{h} 
< 5$ GeV/c is observed for direct photon triggers in Au+Au as 
compared to $p$+$p$.  Furthermore, the level of suppression is 
found to be consistent with the single particle suppression rate 
and the importance of energy-loss geometry, notably the expectation 
of surface emission in the kinematic range sampled.  
A possible indication that energy-loss geometry may also be important in 
dijet suppression is that $\gamma$-$h$ suppression $I_{\rm AA}$ is also 
observed to be quite similar to that of dihadron suppression in central 
events; however, the current precision of the data does not exclude 
substantial differences.
In the $p$+$p$ data $z_{\rm T}$ scaling 
is observed, suggesting that the measured $z_{\rm T}$ distribution 
(Fig.~\ref{fig:xE}) is a statistically acceptable representation 
of the fragmentation function of the quark jet recoiling away from 
the direct photon.  Improvement of the statistical and systematic 
precision of the measurements should allow further tests of vacuum 
fragmentation expectations in p+p collisions and insights into 
details of the medium modification of jet fragmentation in Au+Au.

\section*{Acknowledgements}   

We thank the staff of the Collider-Accelerator and Physics
Departments at Brookhaven National Laboratory and the staff of
the other PHENIX participating institutions for their vital
contributions.  We acknowledge support from the 
Office of Nuclear Physics in the
Office of Science of the Department of Energy,
the National Science Foundation, 
a sponsored research grant from Renaissance Technologies LLC, 
Abilene Christian University Research Council, 
Research Foundation of SUNY, 
and Dean of the College of Arts and Sciences, Vanderbilt University (U.S.A),
Ministry of Education, Culture, Sports, Science, and Technology
and the Japan Society for the Promotion of Science (Japan),
Conselho Nacional de Desenvolvimento Cient\'{\i}fico e
Tecnol{\'o}gico and Funda\c c{\~a}o de Amparo {\`a} Pesquisa do
Estado de S{\~a}o Paulo (Brazil),
Natural Science Foundation of China (People's Republic of China),
Ministry of Education, Youth and Sports (Czech Republic),
Centre National de la Recherche Scientifique, Commissariat
{\`a} l'{\'E}nergie Atomique, and Institut National de Physique
Nucl{\'e}aire et de Physique des Particules (France),
Ministry of Industry, Science and Tekhnologies,
Bundesministerium f\"ur Bildung und Forschung, Deutscher
Akademischer Austausch Dienst, and Alexander von Humboldt Stiftung (Germany),
Hungarian National Science Fund, OTKA (Hungary), 
Department of Atomic Energy (India), 
Israel Science Foundation (Israel), 
Korea Research Foundation and Korea Science and Engineering Foundation (Korea),
Ministry of Education and Science, Rassia Academy of Sciences,
Federal Agency of Atomic Energy (Russia),
VR and the Wallenberg Foundation (Sweden), 
the U.S. Civilian Research and Development Foundation for the
Independent States of the Former Soviet Union, 
the US-Hungarian Fulbight Foundation for Educational Exchange,
and the US-Israel Binational Science Foundation.

\def\IJMPA{{Int. J. Mod. Phys.}~{\bf A}}
\def\JPG{{J. Phys}~{\bf G}}
\def\NCA{Nuovo Cimento}
\def\NIM{Nucl. Instrum. Methods}
\def\NIMA{{Nucl. Instrum. Methods}~{\bf A}}
\def\NPA{{Nucl. Phys.}~{\bf A}}
\def\NPB{{Nucl. Phys.}~{\bf B}}
\def\PLB{Phys. Lett. B}
\def\PLC{Phys. Repts.\ }
\def\PRL{Phys. Rev. Lett.\ }
\def\PRD{Phys. Rev. D}
\def\PRC{Phys. Rev. C}
\def\ZPC{{Z. Phys.}~{\bf C}}
\def\etal{{\it et al.}}



\end{document}